\pdfoutput=1
\documentclass[letterpaper,twocolumn,10pt]{article}
\usepackage{style}
\usepackage[T1]{fontenc}
\usepackage[utf8]{inputenc}
\usepackage{times}

\usepackage{tikz}
\usepackage{amsmath,amssymb,amsfonts}
\usepackage{algorithmic}
\usepackage{graphicx}
\usepackage{textcomp}
\usepackage{xcolor}
\usetikzlibrary{intersections}
\usepackage{pgfplots}
\usepgfplotslibrary{fillbetween}
\usepgfplotslibrary{groupplots}
\pgfplotsset{compat=1.18}
\usepackage{subcaption} 
\usepackage{pifont}
\usepackage{algorithm}
\usepackage{listings}
\usepackage{multirow}
\begin{document}

\date{}

\title{HOCCL: Offloading Collective Communication from GPU Cores to Accelerate Distributed Training \\
}

\author{
{\small\normalfont Yao Fei$^{1}$, Gongming Zhao$^{1}$, Hongli Xu$^{1}$,
Jin Fang$^{1}$, Jiacheng Zhu$^{1}$}\\[-1pt]
{\small\normalfont Shuo Xu$^{2}$, Kun Huang$^{3}$, Zhuolong Yu$^{1}$}\\[2pt]
{\footnotesize\normalfont $^{1}$University of Science and Technology of China, China}\\[-1pt]
{\footnotesize\normalfont $^{2}$China Mobile (Suzhou) Software Technology Co., Ltd., China
\quad $^{3}$Pengcheng Laboratory, China}
}

\maketitle

\begingroup
\renewcommand{\thefootnote}{}
\footnotetext{This work has been submitted to IEEE for possible publication. Copyright may be transferred without notice, after which this version may no longer be accessible.}
\addtocounter{footnote}{-1}
\endgroup

\begin{abstract}

Large language model training involves massive computation on GPU streaming multiprocessors (SMs), the primary compute units of GPUs. Since SMs host specialized accelerators such as Tensor Cores, their efficient utilization is critical to training efficiency.  Unfortunately, existing collective communication systems compete with computation for SMs, as they consume SMs for communication-related data movement and synchronization operations.

We observe that communication can, in principle, be driven by DMA engines, thereby eliminating SM involvement in communication. Based on this insight, we propose HOCCL, a zero-SM collective communication framework consisting of three components: a stream manager, a point-to-point (P2P) executor, and a collective scheduler. The stream manager preserves operator-level temporal ordering with other GPU kernels. The P2P executor enables zero-SM point-to-point communication, while the collective scheduler orchestrates P2P transfers to maximize bandwidth. Experiments show that HOCCL preserves near-peak communication performance, achieving within 3\% of the state of the art on average, while eliminating communication occupancy on nearly 10\% of total GPU SMs. By freeing SM resources for computation, HOCCL improves end-to-end training throughput by up to 5\%.

\end{abstract}
\section{Introduction}
\label{section:intro}

In recent years, large language models (LLMs) have achieved remarkable success and gained widespread adoption, demonstrating expert-level performance across a variety of domains such as natural language understanding \cite{achiam2023gpt,anil2023palm2} and code generation \cite{li2022competition}. Training large language models involves massive computation \cite{hoffmann2022chinchilla,chen2021codex,li2022alphacode}, which is executed primarily on streaming multiprocessors (SMs), the primary execution units of GPUs \cite{nvidia2026cuda_programming_guide}. Specifically, SMs integrate highly specialized hardware (e.g., Tensor Cores \cite{nvidia2020a100_architecture}) to provide sufficient computation capacity for deep learning.  Therefore, fully utilizing SMs is critical to training efficiency. 

\begin{table}[t]
\centering
\small
\setlength{\tabcolsep}{6pt}
\renewcommand{\arraystretch}{1.12}
\begin{tabular}{llcccc}
\hline
Model & SM Usage & TP & PP & DP & EP \\
\hline
\multirow{2}{*}{GPT-6.7B}
  & Peak & 16 & 8 & 8 & -- \\
  & Avg. & 4.8 & 7.1 & 4.1 & -- \\
\hline
\multirow{2}{*}{DeepSeek-V2-Lite}
  & Peak & -- & 8 & 8 & 32 \\
  & Avg. & -- & 7.4 & 2.8 & 9.8 \\
\hline
\end{tabular}
\caption{Peak and time-averaged communication SM occupancy under different parallelization strategies during training on 32 H800 GPUs, each with 132 SMs.}
\label{tab:model_traffic_demo}
\end{table}


However, our measurements uncover an underappreciated inefficiency in traditional training systems: communication consumes a non-negligible number of SMs that would otherwise be available to computation. On a platform with 32 H800 GPUs (each H800 has 132 SMs), we use widely adopted parallelization strategies \cite{shoeybi2019megatron} to train two representative workloads: a GPT-style 6.7B dense model and an MoE model, DeepSeek-V2-Lite \cite{deepseekv2}. Using PyTorch Profiler \cite{pytorch_profiler_recipe}, we measure the runtime overhead of communication kernels and report their average/peak SM usage in Table~\ref{tab:model_traffic_demo}. Although TP and EP exhibit the highest peak SM usage, with peaks of 16 and 32 SMs, respectively, they overlap little with computation and therefore do not contend with computation for SM resources. In contrast, PP and DP are explicitly designed to overlap with computation \cite{zhao2023fsdp,huang2019gpipe,narayanan2019pipedream}, so their SM occupancy correspondingly reduces the SM capacity available to compute kernels. During GPT-6.7B training, PP and DP together occupy an average of 11.2 (7.1 + 4.1) SMs over time, accounting for nearly 10\% of the total GPU SMs on an H800. This level of SM occupancy can reduce the performance of compute kernels by more than 10\% \cite{elvinger2025gpuinterference,lu2025conco,bakita2025hardware} and will translate into millions of dollars of additional training cost \cite{cottier2024rising}.

\begin{figure*}[t]
\centering

\begin{subfigure}[t]{0.33\textwidth}
\centering
\includegraphics[width=\linewidth]{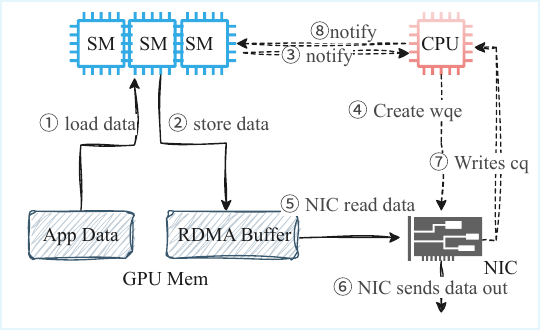}
\subcaption{NCCL}
\label{fig:ibrc}
\end{subfigure}
\begin{subfigure}[t]{0.33\textwidth}
\centering
\includegraphics[width=\linewidth]{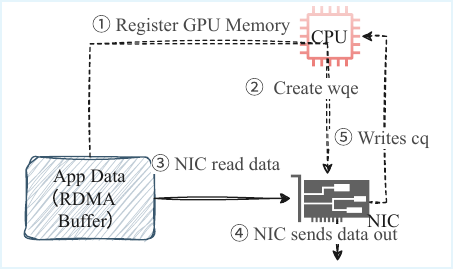}
\subcaption{VCCL}
\label{fig:vccl}
\end{subfigure}
\begin{subfigure}[t]{0.33\textwidth}
\centering
\includegraphics[width=\linewidth]{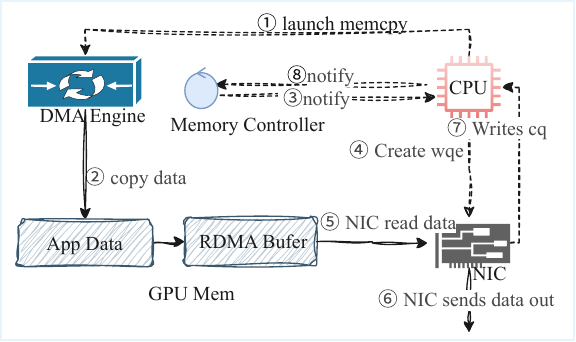}
\subcaption{HOCCL}
\label{fig:ibcpu}
\end{subfigure}

\caption{Data-plane and control-plane differences among NCCL, VCCL, and HOCCL. Solid lines denote the data path, and dashed lines denote the control path. NCCL is a fully SM-centric communication mechanism. VCCL allows the NIC to directly access user data by registering user buffers as NIC-accessible memory regions. In contrast, HOCCL offloads communication tasks to DMA engines and other memory-operation hardware, thereby reducing the work performed by SMs.}
\label{fig:ibrc_ibcpu_vccl}
\end{figure*}

This naturally raises the question of whether computation-overlapped communication, such as PP and DP, can be implemented without SMs. Existing collective communication systems rely on SMs primarily because an RDMA-capable NIC (RNIC) cannot directly perform RDMA reads or writes on arbitrary user buffers \cite{nvidia_rdma_key_concepts}. Instead, data must first be placed in RDMA-registered memory accessible to the RNIC, and this data movement is typically carried out by SMs \cite{nccl} (see details in Figure \ref{fig:ibrc}). Therefore, a major challenge is how to eliminate the need for SMs to move data into RDMA-registered buffers.

One natural way to bypass this issue, as exemplified by VCCL \cite{chen2025efficient} (see details in Figure \ref{fig:vccl}), is to register user buffers with the RNIC at runtime, allowing the RNIC to access them directly without SMs. In practice, however, this approach faces important practical limitations. First, this approach is incompatible with advanced virtual memory management (VMM) mechanisms (e.g., PyTorch expandable segments \cite{pytorch_cuda_semantics}), which may map a contiguous virtual address range onto non-contiguous GPU physical memory. Buffers managed by such VMM mechanisms cannot be registered with the RNIC, forcing VCCL to forgo efficient memory management and potentially incurring more than 10\% additional fragmentation overhead \cite{huang2025stalloc}. Second, in modern training frameworks such as Megatron-LM \cite{shoeybi2019megatron}, frequently communicated tensors (e.g., activations, gradients) are dynamically allocated and freed at runtime \cite{megatron_lm_github} to save memory, so their addresses change unpredictably. This makes repeated RDMA buffer registration unavoidable. In practice, the overhead averages several milliseconds but can fluctuate significantly, sometimes reaching tens of milliseconds \cite{si2025collective}. In our experiments, this overhead will lead to nearly 10\% end-to-end training slowdown (see details in Section \ref{section:e2e}).

We therefore turn to a more practical solution: offloading data movement from SMs to dedicated DMA (Direct Memory Access) engines. DMA engines are CPU-controlled hardware units for bulk GPU memory movement. In current training systems, DMA engines are typically used only for data loading at the beginning of training and result offloading at the end, leaving them lightly utilized during most of training. This suggests that repurposing DMA engines for communication is feasible from a resource-availability perspective. Doing so efficiently, however, is nontrivial and raises three key challenges. First, it introduces a \textit{CPU–GPU coordination challenge}. DMA engines are CPU-driven, but the CPU does not directly observe the execution state needed for timely control decisions. In practice, the GPU remains the primary execution substrate, and the progress of both DMA operations and compute kernels is primarily reflected in GPU-side state. Efficiently coordinating the control plane between the CPU and GPU is therefore nontrivial. Second, it introduces a \textit{transfer latency challenge}. Compared with SM-based copies, DMA engines incur much higher startup latency due to their hardware characteristics and the long control path from the CPU to the DMA engine. In our measurements, this startup latency is about 5$\times$ that of SM-based copying, which leads to more than 20\% communication performance loss. Therefore, effectively hiding this latency is critical. Third, it introduces a \textit{communication scheduling challenge}. Unlike SMs, DMA engines are fixed-function hardware and cannot process multiple transfer tasks in parallel. This forces multiple communication flows to share DMA service in a serialized manner, making collective communication scheduling particularly challenging: an improper DMA task order can lead to deadlock, while poor DMA task scheduling can cause bandwidth imbalance and long-tail latency.

To overcome these challenges, we present HOCCL, a zero-SM collective communication system. HOCCL has three key components. The \textit{stream manager} builds a reliable and efficient synchronization mechanism through handshake-based coordination between the CPU and GPU, leveraging GPU memory-operation hardware. The \textit{point-to-point executor} employs a multi-stage pipelined transfer mechanism to effectively amortize DMA startup overhead. The \textit{collective scheduler} maximizes bandwidth utilization through topology-aware flow scheduling and flow-size-aware chunk scheduling, while supporting arbitrary user-defined collective communication patterns. The main contributions of this paper are as follows:

\begin{itemize}
\item Through an analysis of execution traces from large-scale model training, we identify a previously underexplored factor that limits training efficiency: communication operations contend with computation for shared compute resources (i.e., SMs), thereby slowing down training and reducing overall throughput.

\item We design and implement HOCCL, an SM-free collective communication framework that combines three components: the \textit{stream manager} provides efficient CPU--GPU synchronization, The \textit{point-to-point executor} employs a multi-stage pipelined design to hide DMA startup latency, and the \textit{collective scheduler} schedules flows and chunks to maximize bandwidth utilization.

\item We evaluate HOCCL through benchmarks and end-to-end Megatron-LM training on a testbed with 32 NVIDIA H800 GPUs. Extensive experiments show that HOCCL preserves near-state-of-the-art communication performance, with less than a 3\% loss compared to NCCL, while using no SMs for communication. In end-to-end training, this removes communication occupancy on nearly 10\% of total GPU SMs and improves training throughput by up to 5\%.
\end{itemize}

\section{Background and Motivation}

\subsection{Distributed Training Workloads}

Training large models typically combines multiple parallelization strategies, such as data parallelism (DP) \cite{li2020pytorch_distributed,rajbhandari2020zero}, pipeline parallelism (PP) \cite{huang2019gpipe,narayanan2019pipedream}, and expert parallelism (EP) \cite{lepikhin2021gshard,fedus2022switch}. These strategies improve scalability by partitioning parameters and activations across GPUs, but they also introduce substantial communication. We classify communication into two categories. We refer to the first as \emph{background traffic}, such as PP and DP traffic, because it can often be overlapped with computation to hide its overhead. We refer to the second as \emph{foreground traffic}, such as TP and EP traffic, because it lies closer to the computation and is harder to hide. 

\textbf{Background Traffic in Distributed Training.} Some communication traffic in distributed training can be viewed as \emph{background traffic}, because it is typically designed to overlap with computation to hide communication overhead. PP and DP are two representative examples. In PP, many schedules \cite{narayanan2019pipedream, li2021chimera,qi2024zerobubble} overlap activation transfers and gradient communication for one micro-batch with the forward or backward computation of another. In DP, prior approaches such as FSDP \cite{zhao2023fsdp} prefetch the parameters of the next layer via AllGather during the computation of the current layer, thereby hiding parameter communication behind computation.

\textbf{Foreground Traffic in Distributed Training.} TP and EP communication occurs within a layer and lies on the critical execution path. Subsequent operators often cannot proceed until collectives such as AllReduce, token dispatch, or token combine complete. As a result, unlike PP or DP traffic, TP/EP communication offers little room for overlap and typically exhibits a more serialized computation--communication pattern. Accordingly, prior ML systems work has focused either on maximizing communication performance \cite{deepep2025,hwang2023tutel} or on reducing execution time through fused computation--communication kernels \cite{punniyamurthy2023fusedcollective,zheng2025tilelink,aimuyo2025flashmoe}.

In this work, we focus on \emph{background traffic}. To quantify how much communication occupies SM resources while compute kernels are active and how this affects computation, we define the \emph{contention rate}, which measures the fraction of SMs occupied by communication while compute kernels are active. Let $t \in \mathcal{T}$ denote time within a training step, where $\mathcal{T}$ is the time interval of that step. Let $C(t)$ be an indicator function that equals 1 when compute kernels are active at time $t$, and 0 otherwise. For a traffic class $x$ (e.g., TP, EP, PP, or DP), we define the \emph{contention rate} $I_x$ as
\begin{equation}
I_x =
\frac{
\int_{\mathcal{T}} C(t)\, SM_x(t)\, dt
}{
N_{\mathrm{SM}} \int_{\mathcal{T}} C(t)\, dt
},
\end{equation}
where $SM_x(t)$ is the number of SMs occupied by traffic class $x$ at time $t$, and $N_{\mathrm{SM}}$ is the total number of SMs on the GPU. We further define $I_{\mathrm{sum}} = \sum_x I_x$ as the total contention rate across all traffic classes.

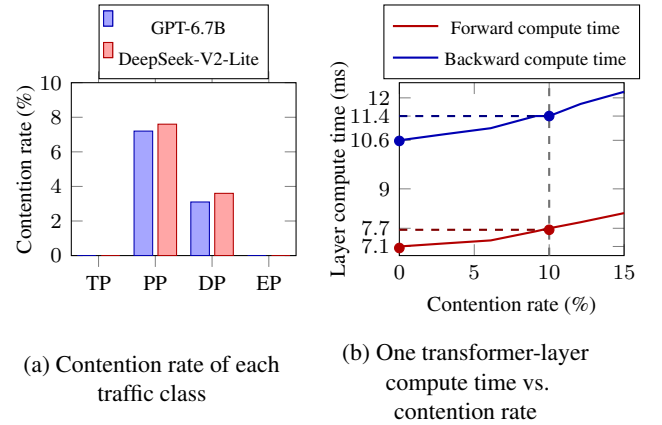
\begin{figure}[t]
\centering
\captionsetup[subfigure]{justification=centering,singlelinecheck=false}
\makebox[\linewidth][c]{%
\begin{subfigure}[t]{0.45\linewidth}
\centering
\vspace{0pt}
\begin{tikzpicture}[baseline=(ax.north)]
\begin{axis}[
  name=ax,
  scale only axis,
  xlabel={\phantom{Contention rate (\%)}},
  ylabel near ticks,
  ylabel={Contention rate (\%)},
  ylabel style={font=\footnotesize,at={(axis description cs:-0.12,0.5)}},
  symbolic x coords={TP,PP,DP,EP},
  xtick=data,
  ymin=0,
  ymax=10,
  ybar,
  bar width=7pt,
  enlarge x limits=0.16,
  width=0.78\linewidth,
  height=0.60\linewidth,
  legend columns=1,
  legend style={
    font=\scriptsize,
    at={(0.5,1.02)},
    anchor=south,
    nodes={text height=1.5ex, text depth=.25ex},
    inner xsep=1.5pt
  },
  legend image code/.code={
    \path[#1] (0cm,0cm) rectangle (0.12cm,0.20cm);
    \draw[#1] (0cm,0cm) rectangle (0.12cm,0.20cm);
  },
  tick label style={font=\footnotesize},
  label style={font=\footnotesize},
  title style={font=\footnotesize},
]
\addplot[fill=blue!35, draw=blue!70!black] coordinates {(TP,0) (PP,7.2) (DP,3.1) (EP,0)};
\addplot[fill=red!35, draw=red!70!black] coordinates {(TP,0) (PP,7.6) (DP,3.6) (EP,0)};
\legend{GPT-6.7B,DeepSeek-V2-Lite}
\end{axis}
\end{tikzpicture}
\caption{Contention rate of each\\ traffic class}
\label{fig:bgtraffic_left}
\end{subfigure}\hspace{0.04\linewidth}%
\begin{subfigure}[t]{0.45\linewidth}
\centering
\vspace{0pt}
\begin{tikzpicture}[baseline=(ax.north)]
\begin{axis}[
  name=ax,
  scale only axis,
  xlabel={Contention rate (\%)},
  ylabel={Layer compute time (ms)},
  ylabel near ticks,
  ylabel style={font=\footnotesize, at={(axis description cs:-0.18,0.5)},},
  xmin=0,
  xmax=15,
  ymin=6.8,
  ymax=12.5,
  xtick={0,5,10,15},
  ytick={7.1,7.7,9,10.6,11.4,12},
  width=0.78\linewidth,
  height=0.60\linewidth,
  legend style={
    font=\scriptsize,
    at={(0.5,1.02)},
    anchor=south,
    nodes={text height=1.5ex, text depth=.25ex}
  },
  tick label style={font=\footnotesize},
  label style={font=\footnotesize},
  title style={font=\footnotesize},
]
\addplot[red!70!black, thick, mark=none] coordinates {
  (0.0,7.1)
  (3.0,7.2)
  (6.1,7.3)
  (9.1,7.6)
  (10,7.7)
  (12.1,7.9)
  (15.0,8.2)
};
\addplot[blue!70!black, thick, mark=none] coordinates {
  (0.0,10.6)
  (3.0,10.8)
  (6.1,11.0)
  (9.1,11.4)
  (10,11.4)
  (12.1,11.8)
  (15.0,12.2)
};
\addplot[black!55, dashed, thick] coordinates {(10,6.8) (10,12.5)};
\addplot[red!50!black, dashed, thick] coordinates {(0.0,7.65) (10,7.65)};
\addplot[blue!50!black, dashed, thick] coordinates {(0.0,11.40) (10,11.40)};
\addplot[red!70!black, only marks, mark=*, mark size=1.8pt] coordinates {
  (0.0,7.05)
  (10,7.65)
};
\addplot[blue!70!black, only marks, mark=*, mark size=1.8pt] coordinates {
  (0.0,10.58)
  (10,11.40)
};
\legend{Forward compute time,Backward compute time}
\end{axis}
\end{tikzpicture}
\caption{One transformer-layer compute time  vs.\\ contention rate}
\label{fig:bgtraffic_right}
\end{subfigure}%
}
\caption{The left figure reports the contention rate $I_x$ of each traffic class during compute. The right figure shows that both forward and backward transformer layer compute time increase noticeably as the contention rate rises.}
\label{fig:bgtraffic}
\end{figure}

Figure.~\ref{fig:bgtraffic_left} reports the contention rate $I_x$ of each parallelization strategy $x$ during training for two representative models.  Since TP and EP communication does not overlap with computation, their contention rates satisfy $I_{\mathrm{TP}} = I_{\mathrm{EP}} = 0$. In contrast, PP and DP account for all contention in both models, causing $I_{\mathrm{sum}}$ to exceed 10\% in each case. Figure.~\ref{fig:bgtraffic_right} further shows the relationship between \emph{contention rate} and the execution time of the main computation in training (i.e., a Transformer layer) based on our experiments. The results show that when $I_{\mathrm{sum}} \approx $ 10\%, forward and backward computation slow down by 8.5\% and 7.7\%, respectively. It also indicates that, for background traffic, even when communication is largely hidden behind computation, it can still impose a non-negligible performance penalty by reducing the SM resources available to compute kernels. 

\subsection{Existing Communication Mechanisms} 
\subsubsection{SM-Based Communication Mechanism}

\textbf{Streaming Multiprocessors (SMs).} SMs are the basic execution units of NVIDIA GPUs and determine their computational capacity. In practice, GPU programs (e.g., general matrix multiplication \cite{nvidia2017cutlass}, Allgather \cite{nvidia2026nccl_collectives}) are issued in the form of kernel functions, which are executed in parallel across these SMs. Each SM integrates several key hardware components, including warp schedulers for instruction dispatch, load/store units for memory operations, CUDA cores and Tensor Cores for computation and fast on-chip memory resources, most notably registers and shared memory. The availability of these remaining resources determines whether an incoming kernel can be scheduled on that SM.

\textbf{SMs' Role in Mainstream GPU Communication Systems.} We summarize NCCL's communication workflow in Figure \ref{fig:ibrc}. The communication kernels in NCCL are responsible both for moving data and for coordinating communication progress and synchronization with the CPU or RNICs. The dashed lines represent how the GPU, the CPU proxy, and the NIC exchange control information, while the solid lines indicate the direction of data movement: \ding{172} SMs reading data from the user buffer into SM-local registers or shared memory, and then \ding{173} writing the data into a staging RDMA buffer. Once the chunk has been staged, GPU threads \ding{174} notify the CPU proxy that the corresponding buffer region is ready. The CPU proxy \ding{175} then prepares NIC work queue entries (WQEs) and rings the doorbell to post the send request. The NIC subsequently \ding{176} reads the staged data from the RDMA buffer and \ding{177} transmits it over the network. After the transfer completes, the NIC \ding{178} writes a completion entry to the completion queue (CQ). The CPU proxy detects the CQ update and notifies the GPU, allowing the communication kernel to continue. 


\subsubsection{Registered-Buffer-Based Communication Mechanism}
VCCL \cite{chen2025efficient} is a representative attempt to reduce SM involvement in communication. Instead of letting GPU communication kernels copy user data into a staging RDMA buffer, it exposes user buffers to the RNIC through runtime registration, so the RNIC can read payloads directly (Figure.~\ref{fig:vccl}). VCCL follows a simpler workflow: \ding{172} the CPU issues a memory region (mr) registration request to the RNIC. After the registration completes, \ding{173} the CPU prepares NIC work queue entries (WQEs) and rings the doorbell to post the send request. The RNIC then \ding{174} reads the data and \ding{175} transmits it. Once the transfer completes, it \ding{176} notifies the CPU of the completion.

This mechanism improves the data path but leaves two practical constraints. The first is memory-layout compatibility: with advanced virtual memory management (VMM), such as PyTorch expandable segments \cite{pytorch_cuda_semantics}, contiguous virtual addresses may correspond to non-contiguous physical pages, which are not directly suitable for RNIC registration. In that case, compatibility with VCCL requires systems to forgo more advanced memory management mechanisms and instead adopt less flexible alternatives, incurring over 10\% memory overhead due to fragmentation \cite{huang2025stalloc}. The second is registration cost. Training stacks such as Megatron-LM \cite{shoeybi2019megatron} frequently allocate and release communication tensors during execution \cite{megatron_lm_github}; their addresses therefore change across iterations, making registration/deregistration a recurring overhead rather than a one-time setup. This overhead is typically at millisecond scale and can spike to tens of milliseconds \cite{si2025collective}, which can lead to up to 10\% end-to-end training slowdown (see details in Section \ref{section:e2e}).

\subsection{DMA Engines}

\textbf{DMA Engines.} A DMA engine (also called a copy engine in NVIDIA documentation) is a dedicated hardware component on the GPU for data movement. DMA engines are controlled by the CPU and can move data without consuming SM resources. They were originally introduced to overlap host-to-device memory transfers with GPU computation on the transferred data. In addition to CPU--GPU transfers, modern DMA engines also support high-bandwidth inter-GPU memory transfers over NVLink, with peak bandwidth reaching up to 900~GB/s \cite{nvidia_gb200_devkit}.

\textbf{DMA Usage in Recent NCCL.} Recent NCCL releases have introduced copy-engine based collectives for intra-node NVLink-domain communication \cite{nccl_ce}. However, this mode is limited to intra-node transfers and does not support cross-node communication over RNIC. Since background communication in distributed training is in most cases cross-node traffic, NCCL's intra-node CE mode does not apply to these scenarios. HOCCL is instead a collective communication system that primarily applies DMA engines to cross-node communication while also supporting intra-node communication.


\section{System Design}
\subsection{Overview of HOCCL}

\begin{figure}[t]
\centering
\includegraphics[width= \columnwidth]{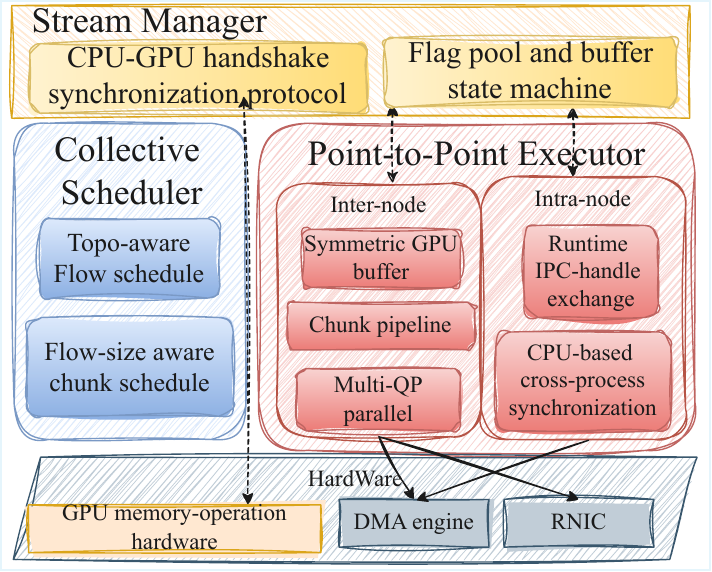}
\caption{System overview of HOCCL. HOCCL consists of three components. The \textit{stream manager} acts as the control hub and coordinates the CPU and GPU using GPU memory-operation hardware. The \textit{point-to-point executor} serves as the main execution engine and handles both intra-node and inter-node communication. The \textit{collective scheduler} serves as the scheduling engine and employs a two-stage algorithm to realize topology-aware collective communication with high bandwidth utilization.}
\label{fig:overview}
\end{figure}


HOCCL addresses three requirements of zero-SM collective communication: (1) preserving CUDA stream ordering, (2) sustaining point-to-point transfer efficiency despite staging and DMA startup overheads, and (3) balancing NVLink, RDMA, DMA-engine utilization. As shown in Figure \ref{fig:overview}, HOCCL consists of three corresponding components: the \textit{stream manager}, the \textit{point-to-point executor}, and the \textit{collective scheduler}.

\textbf{Stream Manager} (Section \ref{sec:ssm}) provides HOCCL's control-plane foundation. The CPU thread acts as the primary control entity, whereas key control information originates on the GPU, including the completion of preceding kernels and the progress state of DMA operations. HOCCL bridges this gap with \textit{GPU memory-operation hardware}, which enables the GPU to directly read from and write to CPU memory. The \textit{handshake synchronization protocol} preserves the ordering between the CPU communication process and the preceding and subsequent dependent kernels on the GPU stream, while the \textit{flag pool and buffer state machine} enable efficient coordination among the RNIC, the CPU proxy, and the DMA engine.

\textbf{Point-to-Point Executor} (Section \ref{sec:p2p}) provides the zero-SM transfer primitive used by higher-level collectives. For RNIC-based inter-node communication, it selects an RNIC-affine \textit{symmetric GPU buffer}, uses \textit{chunk pipelining} to amortize staging overhead, and adopts \textit{multi-QP parallelism} to mitigate DMA startup latency. For intra-node communication, it uses a CPU-based \textit{runtime IPC-handle exchange mechanism} and a \textit{CPU-based cross-process synchronization} mechanism to enable direct DMA-based transfers.

\textbf{Collective Scheduler} (Section \ref{sec:sch}) orchestrates point-to-point flows into efficient collective communication. First, the \textit{Topo-aware flow schedule} decomposes a collective operation into topology-aware stages to hide the intra-node bandwidth overhead introduced by HOCCL's additional DMA copies. Then, the \textit{flow-size-aware chunk schedule} allocates DMA-engine time slices across flows, allowing them to share the same DMA engine and avoid long-tail effects.


Traditional communication libraries launch communication as SM-resident kernels on GPU streams \cite{nvidia2026nccl_streams}. Once enqueued on a stream, these kernels inherit the stream's synchronization semantics, which naturally enforce the correct ordering between communication and the compute kernels before and after it. HOCCL, by contrast, removes such kernels from SMs, making conventional kernel-to-kernel synchronization inapplicable. The \textit{stream manager} must therefore preserve the ordering between communication and surrounding GPU kernels, while also enabling efficient CPU--GPU control exchange for DMA and RNIC coordination. To achieve this, the \textit{stream manager} leverages low-level CUDA driver primitives (e.g., \texttt{cuStreamWriteValue32} and \texttt{cuStreamWaitValue32}) to operate GPU memory-operation hardware, which can exchange control messages with CPU memory without occupying SMs. Built on top of these primitives, the \textit{stream manager} implements an explicit CPU--GPU handshake synchronization protocol.


\subsection{Stream Manager} 
\label{sec:ssm}
\begin{figure}[t]
    \centering
    \includegraphics[width=\linewidth]{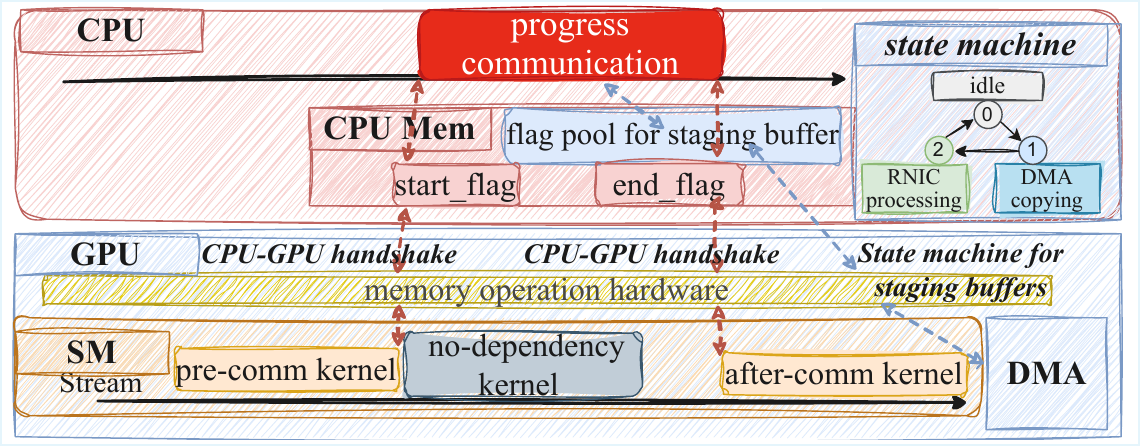}
    \caption{Operation of the stream manager. It leverages memory-operation hardware for low-latency CPU--GPU memory reads and writes, together with a flag pool in CPU memory, to exchange control information.}
    \label{fig:ssm}
\end{figure}

\textbf{CPU--GPU handshake synchronization protocol.} As shown in Fig.~\ref{fig:ssm}, the CPU needs a start notification before initiating communication and must later propagate completion back to the GPU stream. The \textit{stream manager} therefore maintains two notification flags in CPU memory for each GPU stream, allowing direct CPU access and GPU access through memory-operation hardware: \texttt{start\_flag} and \texttt{finish\_flag}. Together, these flags form an explicit CPU--GPU handshake that ensures both sides observe every state transition correctly.
\begin{lstlisting}[language=C++, caption={Handshake synchronization mechanism}, label={lst:example},frame=single, basicstyle=\ttfamily\small,
  columns=fullflexible,
  keepspaces=true]
// On the main thread
// stream1 notifies the CPU to start
cuStreamWriteValue32(stream1,start_flag
,START_VALUE);
cuStreamWaitValue32(stream1,start_flag,0);
// stream1 waits for the CPU until completion
cuStreamWaitValue32(stream1,end_flag
,FINISH_VALUE);
cuStreamWriteValue32(stream1,end_flag,0);

// On the CPU proxy thread
while(start_flag != START_VALUE){}
start_flag = 0;
// Process DMA and RNIC tasks until completion
end_flag = FINISH_VALUE;
while(end_flag != 0){};
\end{lstlisting}

As shown in Listing~\ref{lst:example}, the protocol consists of a start path and a completion path, through which the CPU effectively takes over coordination of the GPU stream. On the start path, the GPU stream writes a request tag (\texttt{START\_VALUE}) to \texttt{start\_flag}. The CPU proxy polls this flag, consumes the request once it becomes visible, and resets it to zero. The GPU stream waits for this reset as an explicit acknowledgement before proceeding. On the completion path, after the communication operation finishes, the CPU writes a completion tag (\texttt{FINISH\_VALUE}) to \texttt{end\_flag} to notify the GPU stream of completion. The GPU stream then resets this flag as the final acknowledgement, completing the end-to-end handshake. This workflow preserves correct stream-ordering semantics without SM-resident communication kernels, while making each state transition explicit and allowing notification state to be safely reused

\textbf{Flag pool and buffer-state coordination.} Once the CPU takes over coordination of the GPU stream and initiates a communication operation, it must manage two asynchronous execution paths simultaneously: RNIC operations on the CPU side and DMA tasks on the GPU side. To coordinate these paths efficiently, HOCCL uses a GPU-CPU-shared flag pool together with a per-chunk buffer state machine. Each staging-buffer chunk is associated with a flag whose value encodes its current state in a three-state machine: \emph{idle}, \emph{DMA copying}, and \emph{RNIC in progress} as shown in Figure \ref{fig:ssm}. These flags are shared between the CPU thread and the GPU stream, allowing both sides to track chunk ownership and execution progress through a unified control structure. 

\begin{figure}[t]
\centering
\begin{subfigure}[t]{\columnwidth}
\centering
\includegraphics[width=\linewidth]{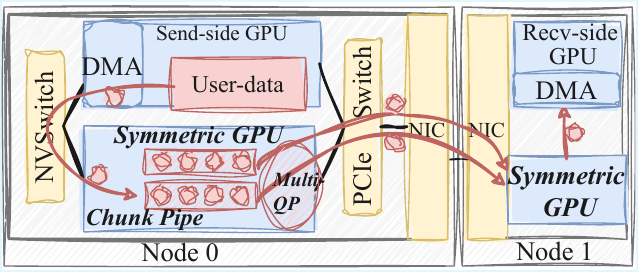}
\subcaption{Detailed design of inter-node communication. The staging buffer is placed on a symmetric GPU and divided across multiple QPs to support multi-stage pipelining.}
\label{fig:inter}
\end{subfigure}

\begin{subfigure}[t]{\columnwidth}
\centering
\includegraphics[width=\linewidth]{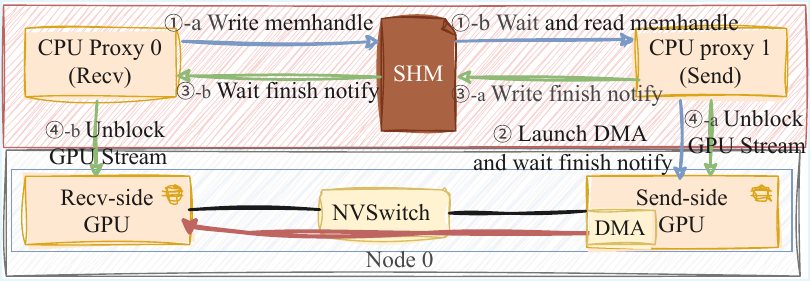}
\subcaption{Detailed design of intra-node communication. HOCCL implements intra-node communication through direct DMA-engine-based transfers, while using the CPU for runtime address exchange and control signaling.}
\label{fig:intranodep2p}
\end{subfigure}
\caption{The upper figure illustrates the data path and pipelined design for inter-node communication, whereas the lower figure shows the CPU--GPU coordination process for intra-node communication.}
\label{fig:intra}
\end{figure}

\subsection{Zero-SM Point-to-Point Executor}
\label{sec:p2p}
\subsubsection{Inter-node Communication}
\textbf{Symmetric GPU buffer.} Inter-node communication must traverse the RNIC, whereas DMA engines support cross-GPU data movement rather than RNIC-mediated transfer. HOCCL therefore allocates two RDMA staging buffers for each GPU on a symmetric peer GPU: one for outgoing data and one for incoming data. Here, a \emph{symmetric GPU} is a GPU under the same PCIe switch and thus at the same distance from the RNIC attached to that switch, as shown in Fig.~\ref{fig:inter}. The DMA engine first moves data over NVSwitch into the staging buffer. Because the local GPU and its symmetric peer are equally distant from the RNIC, the subsequent RDMA transfer incurs no additional path overhead.

\textbf{DMA--RNIC chunk pipelining.} The symmetric-buffer path adds a staging copy. To hide this cost, HOCCL overlaps GPU-side DMA transfers with NIC-side RDMA transfers. It organizes the staging region as a buffer pool, so different chunks can occupy different pipeline stages at the same time. On the sender side, the RNIC transmits ready staging buffers while the DMA engine fills free buffers with subsequent chunks. On the receiver side, the RNIC writes incoming data into receive buffers while the DMA engine drains completed buffers into the final user buffer. This decoupling amortizes DMA startup overhead and preserves high end-to-end transfer efficiency. Figure~\ref{fig:inter} illustrates this design, where the arrows show the transmission path of each chunk.

\textbf{Multi-QP parallelism.} Chunk pipelining overlaps DMA and RNIC work, but a single serial chunk stream would still expose CPU--GPU control latency and DMA-engine launch overhead. HOCCL therefore uses multiple RDMA queue pairs (QPs) per connection, as shown in Figure~\ref{fig:inter}. A QP encapsulates a pair of send and receive work queues together with completion signaling. By striping chunks across multiple QPs (e.g., two or four), HOCCL keeps multiple chunks in flight between each communicating pair. This deeper pipeline hides CPU--GPU control latency and DMA startup overhead behind in-flight chunks and improves RNIC bandwidth utilization.

\subsubsection{Intra-node Communication}

Figure~\ref{fig:intranodep2p} shows HOCCL's intra-node design. HOCCL enables direct send/recv between GPU buffers while resolving the runtime metadata dependencies required for correct communication. Before issuing a DMA write, the sender must obtain the receiver's destination address; before consuming the data, the receiver must know that the transfer has completed. HOCCL handles both metadata exchange and completion notification through CPU-side inter-process messaging, avoiding both staging-buffer-based communication \cite{nccl} and designs that require addresses to be known before communication begins \cite{qin2025mooncake}. As a result, the data path remains direct, asynchronous, and free of SM involvement, enabled by a \textit{runtime IPC-handle exchange mechanism} and a \textit{CPU-based cross-process synchronization mechanism}.

\begin{figure*}[t]
\centering
\includegraphics[width=2\columnwidth]{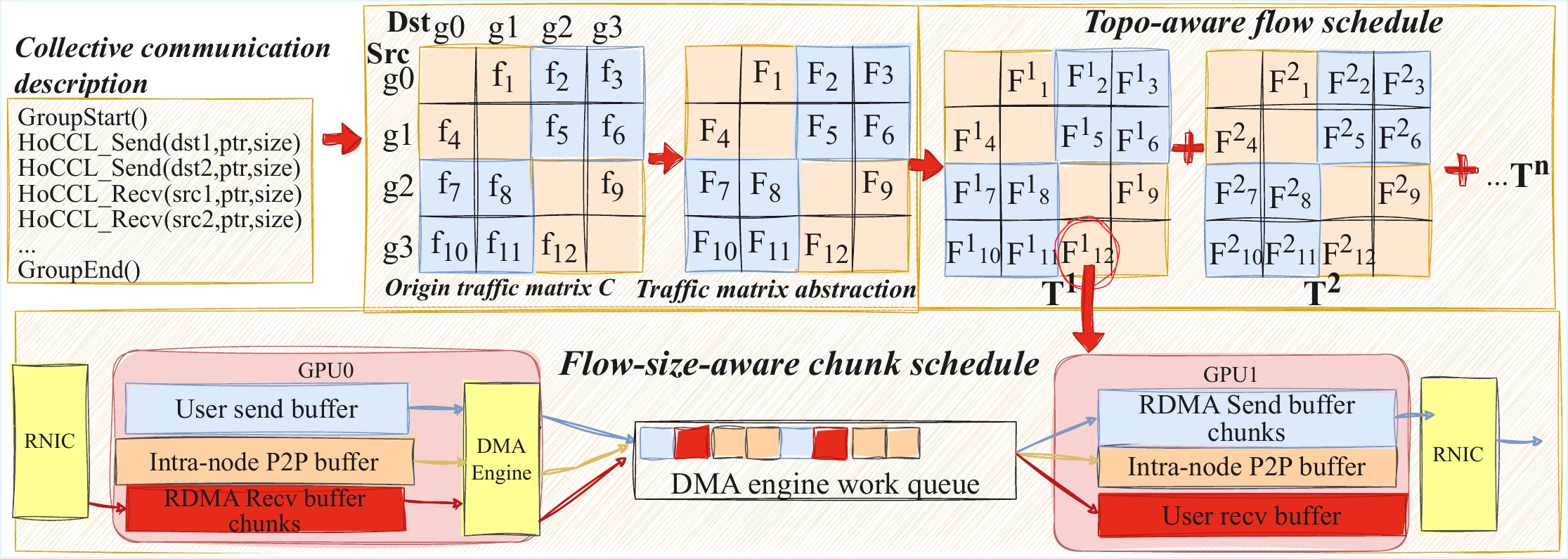}
\caption{Collective scheduler of HOCCL. HOCCL converts a collective description into the origin traffic matrix $C$, expands $C$ into the HOCCL traffic matrix $T$ by adding forwarding traffic, decomposes $T$ into topology-aware stage matrices $T^1,\dots,T^n$, and finally slices the selected flows into chunks for the DMA-engine work queue.}
\label{fig:matrix}
\end{figure*}

\textbf{Runtime IPC-handle exchange.} HOCCL uses POSIX semaphores \cite{opengroup2018posix_semaphore} and shared memory to exchange the control metadata required by the direct path. For each unidirectional $(\textit{rank}, \textit{peer})$ pair, the receiver-side proxy exports the destination buffer as a \texttt{cudaIpcMemHandle} and publishes the handle in shared memory. The sender-side proxy opens the handle at runtime, resolves the receiver's GPU address, and issues a DMA transfer directly into the receiver buffer. Because this exchange is performed by CPU proxy threads, a local send/recv call does not block the calling CPU thread while waiting for the peer to post the matching operation; instead, the proxy completes the exchange once the peer-side metadata becomes available.

\textbf{CPU-based cross-process synchronization.} After the sender-side proxy observes that the DMA write has completed, it posts the completion semaphore associated with the receiver. The receiver-side proxy waits on this semaphore and marks the receive task as finished once the signal arrives. The stream manager can then notify the corresponding GPU stream that the receive operation has completed, allowing subsequent computation on that stream to proceed.

\subsection{Collective Scheduler}
\label{sec:sch}

Figure~\ref{fig:matrix} illustrates the workflow of HOCCL’s collective scheduler. It first translates a user-level collective description into an origin traffic matrix $C$, whose entries are the logical flows $f_1,\dots,f_{12}$. It then expands $C$ into the HOCCL traffic matrix $T$, whose entries $F_1,\dots,F_{12}$ include the forwarding traffic introduced by HOCCL's zero-SM inter-node path. Finally, it decomposes $T$ into stage matrices $T^1,T^2,\dots,T^n$ and schedules the selected flows as chunks on each DMA-engine work queue.

\textbf{Traffic matrix abstraction.} The origin traffic matrix $C\in\mathbb{F}^{G\times G}$ represents the collective exactly as requested by the user, where $G$ is the number of GPUs. Each non-empty entry $C[u,v]$ is a logical flow from source GPU $g_u$ to destination GPU $g_v$. In Figure~\ref{fig:matrix}, for example, $C[0,1]=f_1$, $C[0,2]=f_2$, and $C[3,2]=f_{12}$. Orange boxes denote send/recv pairs within the NVLink domain, which communicate over the intra-node high-bandwidth NVLink fabric( i.e., the intra-node path described above). Blue boxes denote send/recv pairs within the RDMA domain, which communicate via the RNIC (i.e., the inter-node path described above).

HOCCL must convert $C$ into the actual traffic seen by the underlying communication substrates. Let $\mathcal{D}_{\mathrm{RDMA}}$ denote the set of inter-node entries and $\mathcal{D}_{\mathrm{NVLink}}$ denote the set of intra-node NVLink entries. For the topology in Figure~\ref{fig:matrix}, each GPU $g_u$ has a \textit{symmetric GPU peer} $\bar{u}$ : $\bar{0}=1$, $\bar{1}=0$, $\bar{2}=3$, and $\bar{3}=2$. An inter-node logical flow $f_m=C[u,v]$ is realized as three pieces:
\begin{equation}
f_m \Rightarrow f_m' + f_m + f_m'',
\end{equation}
where $f_m'$ is the sender-side NVLink forwarding copy on $(u,\bar{u})$, $f_m$ is the RDMA transfer, and $f_m''$ is the receiver-side NVLink forwarding copy on $(\bar{v},v)$.

Let
\begin{equation}
A_u=\sum_{m\in O_u} f_m',\qquad
B_v=\sum_{m\in I_v} f_m'',
\end{equation}
where $O_u$ is the set of inter-node flows sent by $g_u$, and $I_v$ is the set of inter-node flows received by $g_v$. For an entry $T[u,v]=F_i$, we have
\begin{equation}
T[u,v]=
\begin{cases}
C[u,v] + A_u + B_v, 
& \substack{(u,v)\in\mathcal{D}_{\mathrm{NVLink}}\\ v=\bar{u},}\\
C[u,v], & \text{otherwise.}
\end{cases}
\end{equation}
This notation follows the second matrix in Figure~\ref{fig:matrix}: each $F_i$ is the effective traffic of the corresponding cell after adding forwarding copies.

For example, cell $1$ corresponds to the NVLink edge from $g_0$ to $g_1$. It carries the native intra-node flow $f_1$, the sender-side forwarding copies for $g_0$'s inter-node flows $f_2$ and $f_3$, and the receiver-side forwarding copies for flows whose final destination is $g_1$, namely $f_8$ and $f_{11}$:
\begin{equation}
F_1 = T[0,1] = f_1 + f_2' + f_3' + f_8'' + f_{11}'' .
\end{equation}
Thus, although $C$ contains only user-visible logical flows, $T$ captures the actual NVLink and RDMA traffic that HOCCL must schedule.

\textbf{Topo-aware flow schedule.} After constructing $T$, HOCCL schedules traffic over two types of communication domains: intra-node NVLink domains and inter-node RDMA domains. In Figure~\ref{fig:matrix}, for example, there are three domains: one NVLink domain containing $g_0$ and $g_1$, one NVLink domain containing $g_2$ and $g_3$, and one RDMA domain connecting the two nodes. Executing all entries of $T$ at once can oversubscribe some endpoints while leaving other domains underutilized. This is especially problematic for HOCCL because an RDMA flow also induces NVLink forwarding traffic. Therefore, HOCCL decomposes $T$ into the stage matrices shown on the right side of Figure~\ref{fig:matrix}:
\begin{equation}
P=\{T^1,T^2,\dots,T^n\}.
\end{equation}
Each stage $T^s$ contains a subset of effective flows that can execute together. The decomposition must satisfy three requirements: completeness, per-stage degree bounds, and cross-domain balance.

\textbf{(1) Completeness.} The staged matrices must cover the full HOCCL traffic matrix:
\begin{equation}
\sum_{s=1}^{n} T^s = T \qquad \text{(element-wise)}.
\end{equation}

\textbf{(2) Per-stage degree constraint.} To reduce RDMA congestion, HOCCL restricts each GPU to at most one outgoing and one incoming RDMA flow in each stage:
\begin{equation}
\begin{aligned}
\sum_{v:\,(u,v)\in\mathcal{D}_{\mathrm{RDMA}}}
\mathbb{I}\!\left[T^s[u,v]>0\right] &\le 1,\quad \forall u,s,\\
\sum_{u:\,(u,v)\in\mathcal{D}_{\mathrm{RDMA}}}
\mathbb{I}\!\left[T^s[u,v]>0\right] &\le 1,\quad \forall v,s.
\end{aligned}
\end{equation}
where $\mathbb{I}[\cdot]$ is the indicator function. This constraint makes the blue RDMA entries in each stage contention-free and prevents a single GPU or NIC from becoming a short-term bottleneck.

\textbf{(3) Cross-domain balance.} Since stages are executed serially, the duration of a stage is determined by the slowest communication domain. Let $B_d$ be the bandwidth of domain $d$, and let $V_d$ denote the set of matrix entries belonging to that domain. We define the workload of domain $d$ in stage $s$ as
\begin{equation}
L_d^s = \max_{(u,v)\in V_d}\mathrm{size}\!\left(T^s[u,v]\right).
\end{equation}
The normalized execution time of domain $d$ is $L_d^s/B_d$, so the stage duration is approximated by the slowest domain:
\begin{equation}
\tau^s = \max_d \frac{L_d^s}{B_d}.
\end{equation}
The scheduler therefore seeks a decomposition that reduces the total stage time:
\begin{equation}
\min_{\{T^s\}_{s=1}^{n}} \sum_{s=1}^{n} \tau^s .
\end{equation}
In Figure~\ref{fig:matrix}, the lower-right matrices $T^1,T^2,\dots,T^n$ illustrate this process: each stage contains a contention-free set of RDMA flows and enough NVLink traffic to use the remaining NVLink slack.

\begin{algorithm}[t]
\caption{Birkhoff-like decomposition for the HOCCL traffic matrix}
\label{alg:traffic_decomposition}
\begin{algorithmic}[1]
\REQUIRE HOCCL traffic matrix $T$ with entries $F_i$
\ENSURE Per-stage matrices $P=\{T^1,\dots,T^n\}$
\STATE Initialize the residual matrix $R\leftarrow T$ and $P\leftarrow\emptyset$.
\STATE Decompose native NVLink traffic in each NVLink domain into matching components.
\WHILE{$R$ still has unscheduled traffic}
\STATE Create an empty stage $T^s$.
\STATE Extract a contention-free RDMA matching from the residual matrix.
\STATE Move the largest removable portion of this matching into $T^s$ and subtract it from $R$.
\STATE Add the corresponding sender-side and receiver-side forwarding copies to $T^s$.
\STATE Fill the remaining NVLink slack with native NVLink matching components.
\STATE Append $T^s$ to $P$; $s\leftarrow s+1$.
\ENDWHILE
\end{algorithmic}
\end{algorithm}


\textbf{Birkhoff-like decomposition algorithm.} Algorithm~\ref{alg:traffic_decomposition} decomposes the HOCCL traffic matrix into a sequence of stages. It first initializes the residual matrix and pre-decomposes native NVLink traffic in each NVLink domain into local matching components (lines~1--2). It then repeatedly constructs a new stage by extracting a contention-free RDMA matching from the residual matrix and removing the largest feasible portion (lines~3--6). For each such stage, the algorithm adds the required forwarding copies and uses the remaining NVLink slack to pack local NVLink matching components when possible (lines~7--8). The completed stage is then appended to the output set (line~9). Overall, the algorithm follows the matching-peeling intuition of Birkhoff-von Neumann decomposition \cite{birkhoff1946three} while accounting for forwarding traffic and heterogeneous domain bandwidths.

\textbf{Flow-size-aware chunk scheduling.} Stage-level scheduling determines which flows are active in a stage, while chunk-level scheduling determines how they share the serialized DMA engine at the bottom of Figure~\ref{fig:matrix}. This layer is necessary because a single DMA engine may serve sender-side forwarding copies, receiver-side forwarding copies, and native intra-node transfers, but can execute only one DMA task at a time. A naive per-flow schedule would let large flows monopolize the engine, causing forwarding backlogs and long-tail stragglers. To avoid this, HOCCL divides each flow into fixed-size chunks and allocates DMA service in proportion to flow size. At the beginning of each stage, the scheduler estimates the byte volume of all flows mapped to the same DMA engine and constructs a weighted interleaving of chunks, so larger flows receive more service while smaller flows are still revisited regularly. As illustrated in Figure~\ref{fig:matrix}, a DMA engine may interleave chunks from native intra-node traffic, sender-side forwarding traffic, and receiver-side forwarding traffic within the same stage. This flow-size-aware policy keeps per-flow service rates aligned with demand, allowing flows in the same stage to complete at similar times.

\section{Limitations}

HOCCL has several limitations. First, HOCCL introduces additional NVLink traffic, because each RDMA transfer is accompanied by sender-side and receiver-side forwarding copies. In principle, this extra traffic may contend with latency-sensitive foreground communication such as TP, since both use the same scale-up fabric. In practice, however, the interference is typically limited. The reason is that HOCCL's extra NVLink traffic is  only serves to feed or drain RDMA transfers, and its sustainable rate is therefore fundamentally bounded by the scale-out bandwidth. Since modern GPU clusters usually provide substantially higher scale-up bandwidth than scale-out bandwidth (900 GB/s vs. 50GB/s), the additional NVLink traffic often consumes only a fraction of the available intra-node bandwidth. To further protect latency-sensitive foreground traffic, HOCCL employs a \textit{switch mechanism} that gives strict priority to foreground communication: whenever TP traffic is active, HOCCL temporarily suspends its background tasks and resumes them only after the foreground traffic completes. This design prevents HOCCL from extending the critical path of foreground communication. Second, HOCCL does not naturally support reduction-based operators such as AllReduce and ReduceScatter. These operators require communication to be tightly coupled with reduction computation, and achieving high performance typically relies on SM-side execution. As a result, such operators fall outside HOCCL's target scope. 
\section{Implementation}
We implement HOCCL by extending the NCCL codebase \cite{nccl}. The core HOCCL runtime consists of more than 3K lines of C++ code. To make HOCCL usable in existing training frameworks, we also build a PyTorch-compatible backend \cite{pytorch2026distributed}. This backend is implemented as a lightweight Python extension with approximately 1K lines of C++ code, and its interface is designed to match PyTorch's \texttt{torch.distributed} backend abstraction. As a result, HOCCL can be registered and invoked in PyTorch with minimal integration effort. In our current prototype, switching between NCCL and HOCCL requires only changing an environment variable.

\section{Evaluation}

Our evaluation tests three claims about HOCCL. First, offloading communication from SMs should preserve high point-to-point and collective bandwidth. Second, the additional CPU control work introduced by host-driven progress should remain off the training critical path. Third, eliminating communication-related SM occupancy should improve end-to-end training throughput. We evaluate these claims with communication microbenchmarks, CPU-overhead measurements, and Megatron-LM training workloads.

\pgfplotsset{
  table/col sep=comma,
  table/trim cells=true
}
\definecolor{ncclblue}{RGB}{76,140,210}
\definecolor{vcclgreen}{RGB}{102,176,120}
\definecolor{hocclred}{RGB}{222,114,114}
\definecolor{ncclone}{RGB}{53,104,180}
\definecolor{nccltwo}{RGB}{66,152,181}
\definecolor{ncclfour}{RGB}{126,87,194}
\definecolor{nccleight}{RGB}{230,126,34}
\definecolor{ncclce}{RGB}{70,70,70}
\definecolor{benchgray}{RGB}{120,120,120}

\subsection{Experiment Setup}

\begin{figure}[t]
\centering
\begin{subfigure}[t]{0.46\linewidth}
\centering\vspace{0pt}
\begin{tikzpicture}
\begin{axis}[
  title={Intra-node},
  xmode=log,
  log ticks with fixed point,
  log basis x=2,
  xlabel={Communication Size (MB)},
  ylabel={Algorithm Bandwidth (GB/s)},
  xtick={2,4,8,16,64,256,1024},
  xticklabels={2,4,8,16,64,256,1024},
  scale only axis,
  width=0.75\linewidth,
  height=0.65\linewidth,
  legend columns=4,
  legend style={font=\tiny, at={(0.5,1.03)}, anchor=south},
  tick label style={font=\footnotesize},
  label style={font=\footnotesize},
  title style={font=\footnotesize},
  legend columns=2,
  legend style={font=\tiny, at={(0.5,1.03)}, anchor=south},
]
\addplot[hocclred, thick, solid, mark=triangle*, mark size=1.6pt] table [x=size, y={Hoccl}] {data/data1.csv};
\addplot[ncclce, thick, dashed, mark=x, mark size=1.8pt] table [x=size, y={NCCLCE}] {data/data1.csv};
\addplot[ncclone, thick, solid, mark=*, mark size=1.5pt] table [x=size, y={1sm}] {data/data1.csv};
\addplot[nccltwo, thick, solid, mark=square*, mark size=1.5pt] table [x=size, y={2sm}] {data/data1.csv};
\addplot[ncclfour, thick, solid, mark=diamond*, mark size=1.6pt] table [x=size, y={4sm}] {data/data1.csv};
\addplot[nccleight, thick, solid, mark=pentagon*, mark size=1.8pt] table [x=size, y={8sm}] {data/data1.csv};
\legend{HOCCL,NCCL-CE,NCCL-1sm,NCCL-2sm,NCCL-4sm,NCCL-8sm}
\end{axis}
\end{tikzpicture}
\caption{Intra-node P2P}
\label{intra-p2p}
\end{subfigure}
\hspace{0.04\linewidth}
\begin{subfigure}[t]{0.46\linewidth}
\centering\vspace{0pt}
\begin{tikzpicture}
\begin{axis}[
  title={Inter-node},
  xmode=log,
  log ticks with fixed point,
  log basis x=2,
  xlabel={Communication Size (MB)},
  ylabel={Algorithm Bandwidth (GB/s)},
  ylabel style={font=\footnotesize,at={(axis description cs:-0.12,0.5)}},
  xtick={2,4,8,16,64,256,1024},
  xticklabels={2,4,8,16,64,256,1024},
  scale only axis,
  width=0.75\linewidth,
  height=0.65\linewidth,
  legend columns=4,
  legend style={font=\tiny, at={(0.5,1.03)}, anchor=south},
  tick label style={font=\footnotesize},
  label style={font=\footnotesize},
  title style={font=\footnotesize},
  legend columns=2,
  legend style={font=\tiny, at={(0.5,1.03)}, anchor=south},
]
\addplot[hocclred, thick, solid, mark=triangle*, mark size=1.6pt] table [x=size, y={Hoccl}] {data/data2.csv};
\addplot[ncclone, thick, solid, mark=*, mark size=1.5pt] table [x=size, y={1sm}] {data/data2.csv};
\addplot[nccltwo, thick, solid, mark=square*, mark size=1.5pt] table [x=size, y={2sm}] {data/data2.csv};
\addplot[ncclfour, thick, solid, mark=diamond*, mark size=1.6pt] table [x=size, y={4sm}] {data/data2.csv};
\addplot[nccleight, thick, solid, mark=pentagon*, mark size=1.8pt] table [x=size, y={8sm}] {data/data2.csv};
\legend{HOCCL,NCCL-1sm,NCCL-2sm,NCCL-4sm,NCCL-8sm}
\end{axis}
\end{tikzpicture}
\caption{Inter-node P2P}
\label{inter-p2p}
\end{subfigure}

\caption{Point-to-point bandwidth.}
\end{figure}
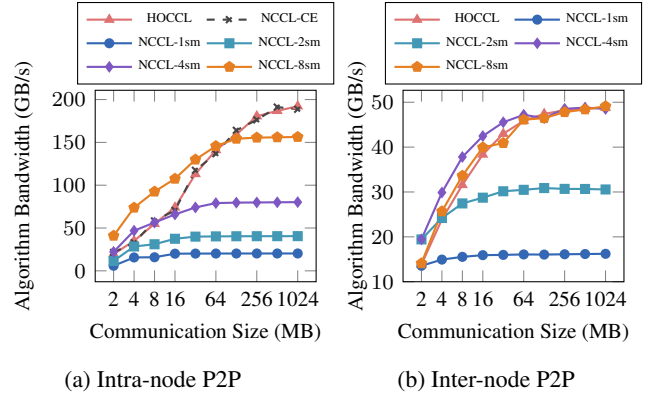

\textbf{Testbed.} We run experiments on an H800 cluster. Each node contains eight NVIDIA H800 80\,GB GPUs. GPUs within a node are connected by NVLink, with up to 400\,GB/s unidirectional bandwidth. Each node also has eight NVIDIA Mellanox ConnectX-5 NICs, each with 50\,GB/s link bandwidth; we use a one-to-one GPU--NIC mapping for inter-node communication. The software stack uses CUDA~12.2, NVIDIA driver~535.161.08, ConnectX-5 firmware~28.38.1902, and NCCL~2.29.7, which was the latest NCCL release available at the time of our experiments \cite{nccl}. We use NVIDIA \texttt{nccl-tests} built against this NCCL version for communication microbenchmarks, and Megatron-LM with a PyTorch-compatible backend for end-to-end training.

\textbf{Baselines.} We compare against NCCL and VCCL. NCCL is the production device-driven baseline, which launches communication kernels on GPU SMs. Since HOCCL's objective is to remove this SM consumption, a comparison only against default NCCL would hide the central resource trade-off. We therefore also evaluate NCCL under explicit communication-kernel SM budgets of 1, 2, 4, and 8 SMs, making the bandwidth--SM trade-off visible. For intra-node experiments covered by NCCL's copy-engine mode, we also report NCCL-CE \cite{nccl_ce}; because this mode does not support cross-node RDMA transfers, it is not included in inter-node P2P or collective benchmarks that cross nodes. VCCL is the most relevant host-driven baseline because it allows the NIC to access user buffers directly after registration.

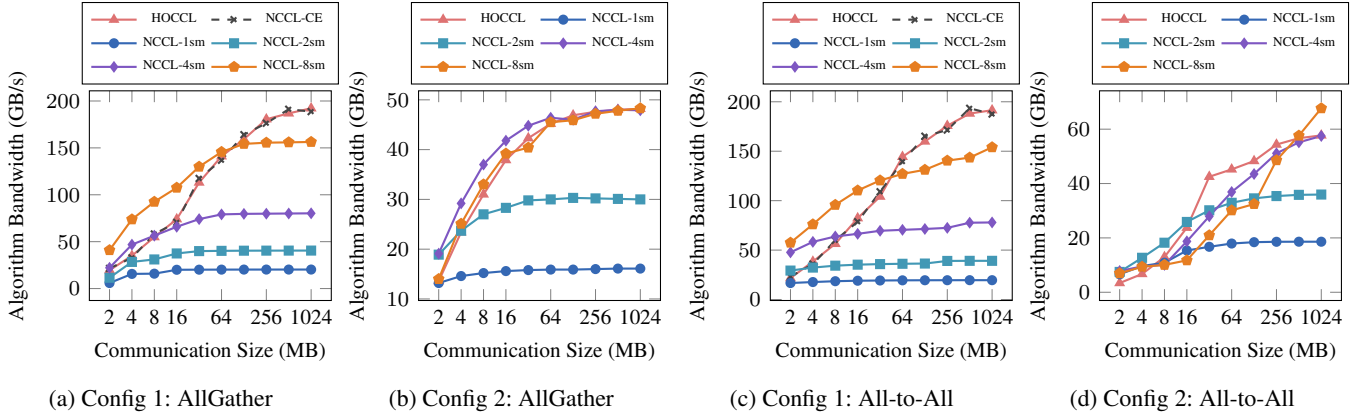
\begin{figure*}[t]
\centering

\begin{subfigure}[t]{0.24\textwidth}
\centering\vspace{0pt}
\begin{tikzpicture}
\begin{axis}[
  xmode=log,
  log ticks with fixed point,
  log basis x=2,
  xlabel={Communication Size (MB)},
  ylabel={Algorithm Bandwidth (GB/s)},
  xtick={2,4,8,16,64,256,1024},
  xticklabels={2,4,8,16,64,256,1024},
  scale only axis,
  width=0.75\linewidth,
  height=0.65\linewidth,
  legend columns=4,
  legend style={font=\tiny, at={(0.5,1.03)}, anchor=south},
  tick label style={font=\footnotesize},
  label style={font=\footnotesize},
  title style={font=\footnotesize},
  legend columns=2,
  legend style={font=\tiny, at={(0.5,1.03)}, anchor=south},
]
\addplot[hocclred, thick, solid, mark=triangle*, mark size=1.6pt] table [x=size, y={Hoccl}] {data/data-allgather8.csv};
\addplot[ncclce, thick, dashed, mark=x, mark size=1.8pt] table [x=size, y={NCCLCE}] {data/data-allgather8.csv};
\addplot[ncclone, thick, solid, mark=*, mark size=1.5pt] table [x=size, y={1sm}] {data/data-allgather8.csv};
\addplot[nccltwo, thick, solid, mark=square*, mark size=1.5pt] table [x=size, y={2sm}] {data/data-allgather8.csv};
\addplot[ncclfour, thick, solid, mark=diamond*, mark size=1.6pt] table [x=size, y={4sm}] {data/data-allgather8.csv};
\addplot[nccleight, thick, solid, mark=pentagon*, mark size=1.8pt] table [x=size, y={8sm}] {data/data-allgather8.csv};
\legend{HOCCL,NCCL-CE,NCCL-1sm,NCCL-2sm,NCCL-4sm,NCCL-8sm}
\end{axis}
\end{tikzpicture}
\caption{Config 1: AllGather}
\label{8allgather}
\end{subfigure}\hfill
\begin{subfigure}[t]{0.24\textwidth}
\centering\vspace{0pt}
\begin{tikzpicture}
\begin{axis}[
  xmode=log,
  log ticks with fixed point,
  log basis x=2,
  xlabel={Communication Size (MB)},
  ylabel={Algorithm Bandwidth (GB/s)},
  xtick={2,4,8,16,64,256,1024},
  xticklabels={2,4,8,16,64,256,1024},
  scale only axis,
  width=0.75\linewidth,
  height=0.65\linewidth,
  legend columns=4,
  legend style={font=\tiny, at={(0.5,1.03)}, anchor=south},
  tick label style={font=\footnotesize},
  label style={font=\footnotesize},
  title style={font=\footnotesize},
  legend columns=2,
  legend style={font=\tiny, at={(0.5,1.03)}, anchor=south},
]
\addplot[hocclred, thick, solid, mark=triangle*, mark size=1.6pt] table [x=size, y={Hoccl}] {data/data-allgather16.csv};
\addplot[ncclone, thick, solid, mark=*, mark size=1.5pt] table [x=size, y={1sm}] {data/data-allgather16.csv};
\addplot[nccltwo, thick, solid, mark=square*, mark size=1.5pt] table [x=size, y={2sm}] {data/data-allgather16.csv};
\addplot[ncclfour, thick, solid, mark=diamond*, mark size=1.6pt] table [x=size, y={4sm}] {data/data-allgather16.csv};
\addplot[nccleight, thick, solid, mark=pentagon*, mark size=1.8pt] table [x=size, y={8sm}] {data/data-allgather16.csv};
\legend{HOCCL,NCCL-1sm,NCCL-2sm,NCCL-4sm,NCCL-8sm}
\end{axis}
\end{tikzpicture}
\caption{Config 2: AllGather}
\label{16allgather}
\end{subfigure}\hfill
\begin{subfigure}[t]{0.24\textwidth}
\centering\vspace{0pt}
\begin{tikzpicture}
\begin{axis}[
  xmode=log,
  log ticks with fixed point,
  log basis x=2,
  xlabel={Communication Size (MB)},
  ylabel={Algorithm Bandwidth (GB/s)},
  xtick={2,4,8,16,64,256,1024},
  xticklabels={2,4,8,16,64,256,1024},
  scale only axis,
  width=0.75\linewidth,
  height=0.65\linewidth,
  legend columns=4,
  legend style={font=\tiny, at={(0.5,1.03)}, anchor=south},
  tick label style={font=\footnotesize},
  label style={font=\footnotesize},
  title style={font=\footnotesize},
  legend columns=2,
  legend style={font=\tiny, at={(0.5,1.03)}, anchor=south},
]
\addplot[hocclred, thick, solid, mark=triangle*, mark size=1.6pt] table [x=size, y={Hoccl}] {data/data-alltoall8.csv};
\addplot[ncclce, thick, dashed, mark=x, mark size=1.8pt] table [x=size, y={NCCLCE}] {data/data-alltoall8.csv};
\addplot[ncclone, thick, solid, mark=*, mark size=1.5pt] table [x=size, y={1sm}] {data/data-alltoall8.csv};
\addplot[nccltwo, thick, solid, mark=square*, mark size=1.5pt] table [x=size, y={2sm}] {data/data-alltoall8.csv};
\addplot[ncclfour, thick, solid, mark=diamond*, mark size=1.6pt] table [x=size, y={4sm}] {data/data-alltoall8.csv};
\addplot[nccleight, thick, solid, mark=pentagon*, mark size=1.8pt] table [x=size, y={8sm}] {data/data-alltoall8.csv};
\legend{HOCCL,NCCL-CE,NCCL-1sm,NCCL-2sm,NCCL-4sm,NCCL-8sm}
\end{axis}
\end{tikzpicture}
\caption{Config 1: All-to-All}
\label{8alltoall}
\end{subfigure}\hfill
\begin{subfigure}[t]{0.24\textwidth}
\centering\vspace{0pt}
\begin{tikzpicture}
\begin{axis}[
  xmode=log,
  log ticks with fixed point,
  log basis x=2,
  xlabel={Communication Size (MB)},
  ylabel={Algorithm Bandwidth (GB/s)},
  xtick={2,4,8,16,64,256,1024},
  xticklabels={2,4,8,16,64,256,1024},
  scale only axis,
  width=0.75\linewidth,
  height=0.65\linewidth,
  legend columns=4,
  legend style={font=\tiny, at={(0.5,1.03)}, anchor=south},
  tick label style={font=\footnotesize},
  label style={font=\footnotesize},
  title style={font=\footnotesize},
  legend columns=2,
  legend style={font=\tiny, at={(0.5,1.03)}, anchor=south},
]
\addplot[hocclred, thick, solid, mark=triangle*, mark size=1.6pt] table [x=size, y={Hoccl}] {data/data-alltoall16.csv};
\addplot[ncclone, thick, solid, mark=*, mark size=1.5pt] table [x=size, y={1sm}] {data/data-alltoall16.csv};
\addplot[nccltwo, thick, solid, mark=square*, mark size=1.5pt] table [x=size, y={2sm}] {data/data-alltoall16.csv};
\addplot[ncclfour, thick, solid, mark=diamond*, mark size=1.6pt] table [x=size, y={4sm}] {data/data-alltoall16.csv};
\addplot[nccleight, thick, solid, mark=pentagon*, mark size=1.8pt] table [x=size, y={8sm}] {data/data-alltoall16.csv};
\legend{HOCCL,NCCL-1sm,NCCL-2sm,NCCL-4sm,NCCL-8sm}
\end{axis}
\end{tikzpicture}
\caption{Config 2: All-to-All}
\label{16alltoall}
\end{subfigure}\hfill

\caption{Collective communication bandwidth.}
\end{figure*}

We do not include VCCL in bandwidth-only microbenchmarks. Standard microbenchmarks repeatedly reuse the same communication buffers, which amortizes registration cost and can substantially overstate the benefit of user-buffer registration.

\textbf{Communication benchmarks.} We use \texttt{nccl-tests} \cite{nvidia_nccl_tests} to isolate communication performance. We evaluate point-to-point transfers and two collectives: AllGather and All-to-All. Intra-node P2P runs within a single node, while inter-node P2P uses one GPU pair across nodes. For each collective, we evaluate two configurations: Config~1 represents intra-node communication, and Config~2 represents collectives involving inter-node traffic. NCCL-CE is included for intra-node P2P and Config~1 collectives, where its copy-engine mode applies; it is omitted for Config~2 because NCCL-CE does not support cross-node transfers. We sweep message sizes from 2\,MB to 1\,GB and report algorithmic bandwidth. For NCCL, we additionally vary the number of SMs assigned to communication kernels.

\begin{table}[t]
\centering
\small
\setlength{\tabcolsep}{3pt}
\renewcommand{\arraystretch}{1.08}
\begin{tabular}{lccccc}
\hline
Model & Layers & Hidden & Heads & Seq. & Parallelism \\
\hline
6.7B & 32 & 4096 & 32 & 2048 & PP=4, DP=8 \\
32B & 60 & 6656 & 52 & 2048 & TP=4, PP=4, DP=2 \\
\hline
\end{tabular}
\caption{End-to-end training model configurations.}
\label{tab:e2e_model_config}
\end{table}

\textbf{End-to-end workloads.} We evaluate training with Megatron-LM on all 32 H800 GPUs using GPT-style decoder-only Transformer models. Table~\ref{tab:e2e_model_config} summarizes the model and parallelism configurations. Both workloads use BF16 mixed precision and sequence length 2048. For throughput experiments, we use global batch sizes 32 and 128 for the 6.7B and 32B workloads, respectively; for the compute-time breakdown, we additionally sweep the micro-batch size over 1 and 2. These workloads exercise different compute--communication balances: the smaller model has less compute per communicated byte, while the larger model provides more opportunity to hide host-side work behind GPU execution. In these end-to-end experiments, HOCCL accelerates PP \texttt{send}/\texttt{recv} and DP \texttt{AllGather}. Communication primitives that require reduction semantics, such as \texttt{ReduceScatter}, fall back to NCCL.

\subsection{Communication Benchmark}
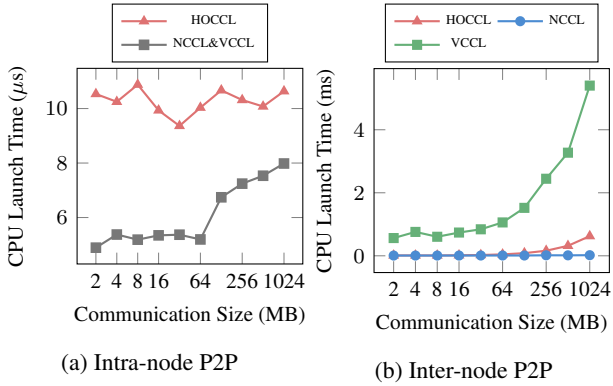
\begin{figure}[t]
\centering

\begin{subfigure}[t]{0.47\linewidth}
\centering\vspace{0pt}
\begin{tikzpicture}
\begin{axis}[
  title={Intra-node},
  xmode=log,
  log ticks with fixed point,
  log basis x=2,
  xlabel={Communication Size (MB)},
  ylabel={CPU Launch Time ($\mu$s)},
  xtick={2,4,8,16,64,256,1024},
  xticklabels={2,4,8,16,64,256,1024},
  scale only axis,
  width=0.75\linewidth,
  height=0.65\linewidth,
  legend style={font=\tiny, at={(0.5,1.03)}, anchor=south},
  tick label style={font=\footnotesize},
  label style={font=\footnotesize},
  title style={font=\footnotesize},
  legend columns=1,
  legend style={font=\tiny, at={(0.5,1.03)}, anchor=south},
]
\addplot[hocclred, thick, solid, mark=triangle*, mark size=1.6pt] table [x=size, y={Hoccl}] {data/data-cpucost8.csv};
\addplot[benchgray, thick, solid, mark=square*, mark size=1.6pt] table [x=size, y={NCCL&VCCL}] {data/data-cpucost8.csv};
\legend{HOCCL,NCCL\&VCCL}
\end{axis}
\end{tikzpicture}
\caption{Intra-node P2P}
\label{cputimeintra}
\end{subfigure}
\begin{subfigure}[t]{0.49\linewidth}
\centering\vspace{0pt}
\begin{tikzpicture}
\begin{axis}[
  title={Inter-node},
  xmode=log,
  log ticks with fixed point,
  log basis x=2,
  xlabel={Communication Size (MB)},
  ylabel={CPU Launch Time (ms)},
  xtick={2,4,8,16,64,256,1024},
  xticklabels={2,4,8,16,64,256,1024},
  scale only axis,
  width=0.75\linewidth,
  height=0.65\linewidth,
  legend columns=4,
  legend style={font=\tiny, at={(0.5,1.03)}, anchor=south},
  tick label style={font=\footnotesize},
  label style={font=\footnotesize},
  title style={font=\footnotesize},
  legend columns=2,
  legend style={font=\tiny, at={(0.5,1.03)}, anchor=south},
]
\addplot[hocclred, thick, solid, mark=triangle*, mark size=1.6pt] table [x=size, y expr=\thisrow{Hoccl}/1000] {data/data-cpucost16.csv};
\addplot[ncclblue, thick, solid, mark=*, mark size=1.5pt] table [x=size, y expr=\thisrow{NCCL}/1000] {data/data-cpucost16.csv};
\addplot[vcclgreen, thick, solid, mark=square*, mark size=1.5pt] table [x=size, y expr=\thisrow{VCCL}/1000] {data/data-cpucost16.csv};
\legend{HOCCL,NCCL,VCCL}
\end{axis}
\end{tikzpicture}
\caption{Inter-node P2P}
\label{cputimeinter}
\end{subfigure}

\caption{CPU-side launch overhead.}
\end{figure}

\textbf{Intra-node P2P.} HOCCL achieves high intra-node bandwidth without communication SMs. As shown in Figure~\ref{intra-p2p}, HOCCL increases from 19.6\,GB/s at 2\,MB to 192.2\,GB/s at 1\,GB. NCCL-CE exhibits similar intra-node performance with HOCCL. NCCL's SM-driven bandwidth, in contrast, scales with the SMs allocated to its communication kernels: under 1/2/4/8 SMs, it plateaus at roughly 20/40/80/156\,GB/s on medium and large messages. NCCL-8SM is faster on small transfers because SM kernels have lower startup latency (41.0 vs.\ 19.6\,GB/s at 2\,MB), but HOCCL catches up at 64--128\,MB and exceeds NCCL-8SM by 23\% at 1\,GB (192.2 vs.\ 156.3\,GB/s). 

\textbf{Inter-node P2P.} HOCCL saturates the inter-node link for large transfers, but pays higher startup cost on small messages. Figure~\ref{inter-p2p} shows that HOCCL scales from 13.8\,GB/s at 2\,MB to 48.8\,GB/s at 1\,GB, close to the effective 50\,GB/s NIC bandwidth. NCCL again depends on the SM budget when constrained: NCCL-1SM and NCCL-2SM plateau around 16 and 31\,GB/s, while 4 SMs are sufficient to nearly saturate the link and 8 SMs add little. HOCCL lags on small messages because DMA startup and chunk-pipeline setup are harder to amortize, but the gap disappears once the transfer becomes bandwidth dominated. Thus, for large inter-node P2P, HOCCL reaches NCCL's best-case bandwidth while avoiding device-side progress.

\textbf{AllGather.} HOCCL preserves strong AllGather bandwidth when NCCL cannot dedicate many SMs to communication. Figures~\ref{8allgather} and~\ref{16allgather} report an 8-GPU intra-node AllGather and a 16-GPU cross-node AllGather. In the 8-GPU intra-node case, NCCL-CE is close to HOCCL across message sizes and reaches 188.7\,GB/s at 1\,GB compared with HOCCL's 192.2\,GB/s. HOCCL is also 2.4$\times$ faster than NCCL-4SM (80.2\,GB/s) at this size, showing that DMA-based intra-node transfers can match NCCL-CE while avoiding communication SMs. The 16-GPU AllGather experiment is conducted on a separate testbed, where we use one GPU from each participating node. This configuration forces all communication onto the NIC/RDMA path rather than the intra-node NVLink path. We choose this setup because AllGather commonly appears in DP communication, and prior large-scale training systems often reserve the fast intra-node fabric for bandwidth-sensitive traffic such as TP while placing DP communication across nodes \cite{zhao2023fsdp,song2023optimuscc,jiang2024megascale}. This setup therefore isolates the inter-node behavior most relevant to our target DP workload. At 1\,GB, HOCCL reaches 48.2\,GB/s, comparable to NCCL-4SM and NCCL-8SM at 47.9 and 48.3\,GB/s, while NCCL-1SM and NCCL-2SM plateau near 16 and 30\,GB/s. NCCL-CE is omitted because it is intra-node only. This result shows that HOCCL extends zero-SM communication to the cross-node AllGather pattern that dominates many DP workloads.

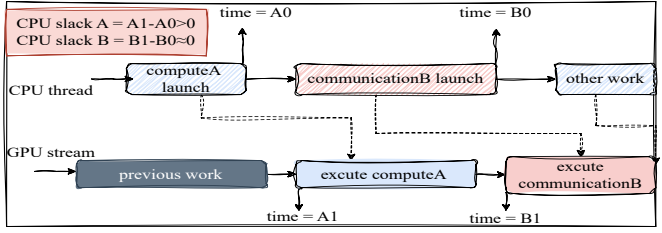
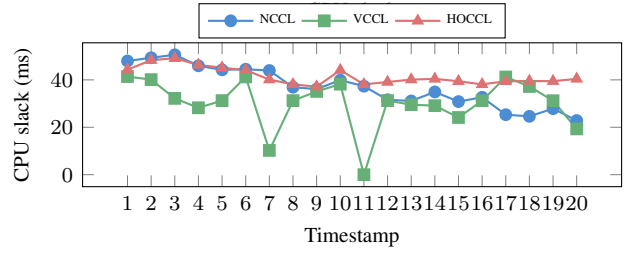
\begin{figure*}[t]
\centering
\begin{subfigure}[t]{0.49\textwidth}
\centering
\includegraphics[width=\linewidth,height=0.35\linewidth]{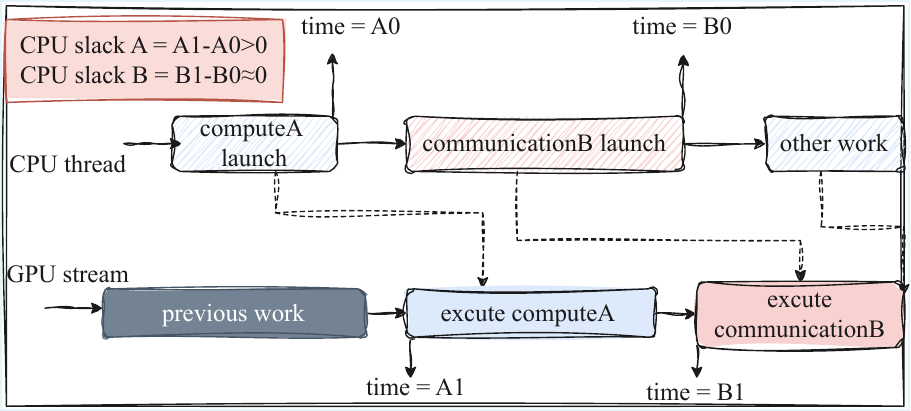}
\subcaption{The definition of CPU slack}
\label{fig:cpugpu_left}
\end{subfigure}%
\hfill
\begin{subfigure}[t]{0.49\textwidth}
\centering
\begin{tikzpicture}
\begin{axis}[
  title={CPU slack},
  xlabel={Timestamp},
  ylabel={CPU slack (ms)},
  xtick=data,
  width=\linewidth,
  height=0.4\linewidth,
  tick label style={font=\footnotesize},
  label style={font=\footnotesize},
  title style={font=\footnotesize},
  legend columns=3,
  legend style={font=\tiny, at={(0.5,1.03)}, anchor=south},
]
\addplot[ncclblue, thick, mark=*] table [x=timestap, y={NCCL}] {data/data-time.csv};
\addplot[vcclgreen, thick, mark=square*] table [x=timestap, y={VCCL}] {data/data-time.csv};
\addplot[hocclred, thick, mark=triangle*] table [x=timestap, y={HOCCL}] {data/data-time.csv};
\legend{NCCL,VCCL,HOCCL}
\end{axis}
\end{tikzpicture}
\subcaption{CPU slack over time}
\label{fig:cpugpu_right}
\end{subfigure}

\caption{CPU slack in asynchronous GPU execution.}
\label{fig:cpugpu}
\end{figure*}

\textbf{All-to-All.} HOCCL is particularly effective for large All-to-All exchanges. On 8 GPUs (Figure~\ref{8alltoall}), HOCCL reaches 191.4\,GB/s at 1\,GB, comparable to NCCL-CE at 187.6\,GB/s, and exceeds NCCL-8SM by 24\% (154.0\,GB/s), while also outperforming all SM-constrained NCCL configurations. At 16 GPUs, bandwidth decreases for both systems because the communication degree and induced intra-node traffic increase. HOCCL reaches 57.7\,GB/s at 1\,GB, matching NCCL-4SM (57.5\,GB/s) and substantially outperforming NCCL-1SM and NCCL-2SM. NCCL-8SM remains faster at 67.7\,GB/s, a 17\% advantage, because it can spend more SM resources to drive more aggressive communication progress. The scalability trend is therefore consistent across collectives: HOCCL is strongest when SMs are scarce or valuable, while NCCL's peak bandwidth can still improve by dedicating more SMs to communication.

\textbf{CPU launch overhead.} HOCCL increases CPU launch cost, but the magnitude depends on whether the path is intra-node or inter-node. Figures~\ref{cputimeintra} and~\ref{cputimeinter} report CPU-side launch overhead for P2P operations. For intra-node P2P, HOCCL costs 9--12\,$\mu$s, compared with 5--8\,$\mu$s for NCCL. The difference comes from HOCCL's CPU-side sender/receiver coordination for direct DMA writes. VCCL is similar to NCCL in this setting because it falls back to NCCL for intra-node transfers rather than providing its own SM-free intra-node path.

For inter-node P2P, NCCL's launch cost remains nearly flat at 8--20\,$\mu$s, while HOCCL grows from 11.8\,$\mu$s at 2\,MB to 624.6\,$\mu$s at 1\,GB. This growth is expected: HOCCL is host-driven and submits more DMA/RDMA work as a transfer is divided into more chunks. VCCL has much higher overhead, from 563\,$\mu$s to 5.4\,ms, because it performs blocking RDMA buffer registration whose cost grows with buffer size.

This overhead result also identifies a practical limitation of user-buffer registration. Under NIC and GPU driver load, RDMA registration latency is highly variable and can spike to 20--30\,ms in our experiments, consistent with prior observations from Meta's large-scale training deployments \cite{si2025collective}. This explains why VCCL can suffer severe jitter despite being host-driven, and motivates HOCCL's use of reusable staging buffers rather than repeatedly registering application tensors.
\subsection{End-to-End Training}
\label{section:e2e}
We next evaluate whether eliminating communication-related SM usage improves full training runs. We integrate HOCCL into Megatron-LM and compare it with NCCL and VCCL on two workloads with different compute--communication balance. We report throughput, per-step compute time, and CPU slack to determine whether HOCCL's host-driven control path affects the critical path of training.
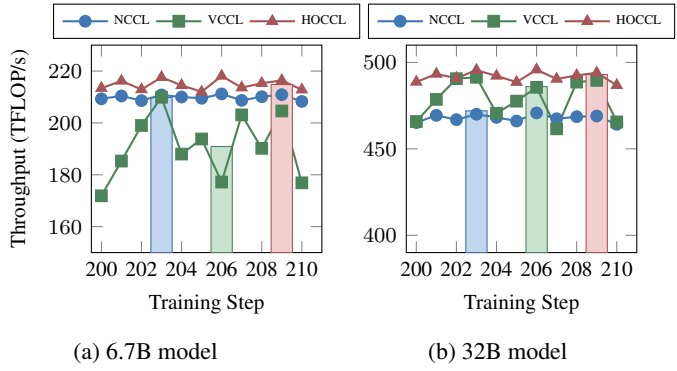
\begin{figure}[t]
\centering

\begin{subfigure}[t]{0.45\linewidth}
\centering\vspace{0pt}
\begin{tikzpicture}
\begin{axis}[
  xlabel={Training Step},
  ylabel={Throughput (TFLOP/s)},
  xmin=199.5,
  xmax=211,
  ymin=150,
  ymax=230,
  xtick={200,202,204,206,208,210,212,214,216},
  xticklabels={200,202,204,206,208,210,,,},
  unbounded coords=jump,
  scale only axis,
  width=0.8\linewidth,
  height=0.72\linewidth,
  tick label style={font=\footnotesize},
  label style={font=\footnotesize},
  title style={font=\footnotesize},
  legend columns=3,
  legend style={font=\tiny, at={(0.5,1.03)}, anchor=south},
]
\addplot[ybar, bar width=8pt, fill=ncclblue!35, draw=ncclblue!80!black, forget plot]
  table [x=x, y=NCCL] {data/e2e-6p5b-avg.csv};
\addplot[ncclblue!85!black, thick, mark=*]
  table [x=step, y=NCCL] {data/e2e-6p5b-steps.csv};
\addplot[ybar, bar width=8pt, fill=vcclgreen!35, draw=vcclgreen!75!black, forget plot]
  table [x=x, y=VCCL] {data/e2e-6p5b-avg.csv};
\addplot[vcclgreen!75!black, thick, mark=square*]
  table [x=step, y=VCCL] {data/e2e-6p5b-steps.csv};
\addplot[hocclred!75!black, thick, mark=triangle*]
  table [x=step, y=HOCCL] {data/e2e-6p5b-steps.csv};
\addplot[ybar, bar width=8pt, fill=hocclred!35, draw=hocclred!75!black, forget plot]
  table [x=x, y=HOCCL] {data/e2e-6p5b-avg.csv};
\legend{NCCL,VCCL,HOCCL}
\end{axis}
\end{tikzpicture}
\caption{6.7B model}
\label{bs32}
\end{subfigure}\hfill
\begin{subfigure}[t]{0.45\linewidth}
\centering\vspace{0pt}
\begin{tikzpicture}
\begin{axis}[
  xlabel={Training Step},
  xmin=199.5,
  xmax=211,
  ymin=390,
  ymax=510,
  xtick={200,202,204,206,208,210,212,214,216},
  xticklabels={200,202,204,206,208,210,,,},
  unbounded coords=jump,
  scale only axis,
  width=0.8\linewidth,
  height=0.72\linewidth,
  tick label style={font=\footnotesize},
  label style={font=\footnotesize},
  title style={font=\footnotesize},
  legend columns=3,
  legend style={font=\tiny, at={(0.5,1.03)}, anchor=south},
]
\addplot[ybar, bar width=8pt, fill=ncclblue!35, draw=ncclblue!80!black, forget plot]
  table [x=x, y=NCCL] {data/e2e-13b-avg.csv};
\addplot[ncclblue!85!black, thick, mark=*]
  table [x=step, y=NCCL] {data/e2e-13b-steps.csv};
\addplot[ybar, bar width=8pt, fill=vcclgreen!35, draw=vcclgreen!75!black, forget plot]
  table [x=x, y=VCCL] {data/e2e-13b-avg.csv};
\addplot[vcclgreen!75!black, thick, mark=square*]
  table [x=step, y=VCCL] {data/e2e-13b-steps.csv};
\addplot[ybar, bar width=8pt, fill=hocclred!35, draw=hocclred!75!black, forget plot]
  table [x=x, y=HOCCL] {data/e2e-13b-avg.csv};
\addplot[hocclred!75!black, thick, mark=triangle*]
  table [x=step, y=HOCCL] {data/e2e-13b-steps.csv};
\legend{NCCL,VCCL,HOCCL}

\end{axis}
\end{tikzpicture}
\caption{32B model}
\label{bs128}
\end{subfigure}

\caption{End-to-end training throughput}
\end{figure}

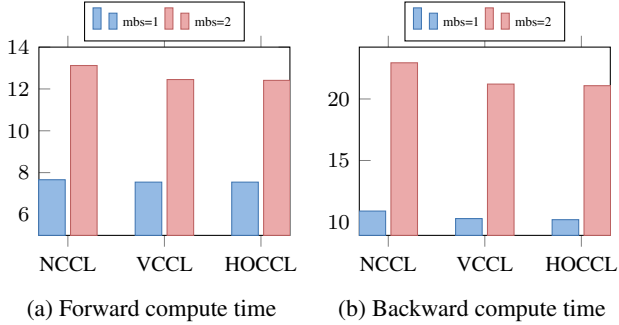
\begin{figure}[t]
\centering

\begin{subfigure}[t]{0.49\linewidth}
\centering\vspace{0pt}
\begin{tikzpicture}
\begin{axis}[
  title={Forward pass},
  xtick=type, 
  xticklabels={NCCL,VCCL,HOCCL}, 
  ybar,
  bar width=10pt,
  enlarge x limits=0.15,
  xtick=data,
  scale only axis,
  ymin=5,
  ymax=14,
  width=0.8\linewidth,
  height=0.6\linewidth,
  legend columns=3,
  legend style={font=\tiny, at={(0.5,1.03)}, anchor=south},
  tick label style={font=\footnotesize},
  label style={font=\footnotesize},
  title style={font=\footnotesize},
]
\addplot[fill=ncclblue!60, draw=ncclblue!85!black] table [x=type, y={size1}] {data/data-fcomp.csv};
\addplot[fill=hocclred!60, draw=hocclred!85!black] table [x=type, y={size4}] {data/data-fcomp.csv};
\legend{mbs=1,mbs=2}
\end{axis}
\end{tikzpicture}
\caption{Forward compute time}
\end{subfigure}
\begin{subfigure}[t]{0.49\linewidth}
\centering\vspace{0pt}
\begin{tikzpicture}
\begin{axis}[
  title={Backward pass},
  xtick=type, 
  xticklabels={NCCL,VCCL,HOCCL},
  ybar,
  bar width=10pt,
  enlarge x limits=0.15,
  xtick=data,
  scale only axis,
  width=0.8\linewidth,
  height=0.6\linewidth,
  legend columns=3,
  legend style={font=\tiny, at={(0.5,1.03)}, anchor=south},
  tick label style={font=\footnotesize},
  label style={font=\footnotesize},
  title style={font=\footnotesize},
]
\addplot[fill=ncclblue!60, draw=ncclblue!85!black] table [x=type, y={size1}] {data/data-bcomp.csv};
\addplot[fill=hocclred!60, draw=hocclred!85!black] table [x=type, y={size4}] {data/data-bcomp.csv};
\legend{mbs=1,mbs=2}
\end{axis}
\end{tikzpicture}
\caption{Backward compute time}
\label{bct}
\end{subfigure}

\caption{Compute time per micro-batch}
\label{perstep}
\end{figure}

\textbf{Training throughput.} HOCCL improves end-to-end throughput because the SMs saved from communication become available to training kernels. Figures~\ref{bs32} and~\ref{bs128} report per-step throughput for the 6.7B and 32B models. On the 6.7B workload, NCCL achieves 209.8\,TFLOP/s on average, VCCL reaches 190.9\,TFLOP/s, and HOCCL reaches 214.8\,TFLOP/s. HOCCL improves over NCCL by 2.4\%, whereas VCCL is 9.0\% slower than NCCL. VCCL also exhibits large jitter, ranging from 171.9 to 209.9\,TFLOP/s, while HOCCL remains between 212.0 and 218.1\,TFLOP/s.

The larger 32B workload amplifies HOCCL's benefit. NCCL averages 467.8\,TFLOP/s, VCCL averages 478.7\,TFLOP/s, and HOCCL averages 491.7\,TFLOP/s. HOCCL therefore improves over NCCL by 5.1\% and over VCCL by 2.7\%. VCCL improves over NCCL on average in this workload, but its throughput still varies from 461.6 to 491.4\,TFLOP/s because registration and host-side latency occasionally enter the critical path. HOCCL remains both faster and more stable, ranging only from 486.8 to 495.8\,TFLOP/s. These results support the central claim that HOCCL converts saved communication SMs into higher training throughput while avoiding VCCL's registration-induced instability.



\textbf{Impact on compute time.} Removing communication kernels from SMs measurably shortens overlapped compute. Figure~\ref{perstep} reports forward and backward compute time for two micro-batch sizes. For the forward pass, HOCCL reduces average compute time from 7.66\,ms to 7.55\,ms at mbs=1 (1.5\%) and from 13.12\,ms to 12.41\,ms at mbs=2 (5.4\%). For the backward pass, HOCCL reduces time from 10.89\,ms to 10.19\,ms at mbs=1 (6.4\%) and from 22.94\,ms to 21.08\,ms at mbs=2 (8.1\%). These reductions directly connect the end-to-end throughput gains to the design goal of freeing SM resources for training kernels.

\textbf{Extra CPU cost analysis.} We next analyze whether the additional CPU time overhead introduced by HOCCL can make the CPU side a training bottleneck, and explain why VCCL exhibits performance fluctuations. Modern training stacks enqueue work asynchronously: while the GPU executes previously issued kernels, the CPU prepares future work. As shown in Figure \ref{fig:cpugpu_left}, we define \emph{CPU slack} as the difference between an operator's GPU start time and the end of its CPU launch (i.e., CPU slack = GPU start time - CPU launch end time). CPU slack A is defined as the gap between the CPU submission time of compute kernel A and its actual execution time on the GPU; a positive value indicates that the CPU is not the bottleneck. In contrast, CPU slack B equals zero, meaning that CPU submission is no longer ahead of the execution of prior GPU work. In this case, excessive CPU overhead becomes a system bottleneck. Using \texttt{torch.profiler}, we sample operators and record their CPU enqueue timestamps and GPU start times. 

Figure~\ref{fig:cpugpu_right} shows that NCCL and HOCCL maintain positive slack throughout the measured window, typically 20--50\,ms. Thus, although HOCCL launches more host-side DMA/RDMA work than NCCL, the extra work is absorbed by existing CPU slack and does not shift the bottleneck from GPU execution to CPU scheduling. VCCL has much higher variance and reaches zero slack at timestamp~11, indicating that blocking registration can stall GPU execution. This explains the throughput jitter observed in Figure~\ref{bs32} and Figure~\ref{bs128}, and highlights a trade-off: HOCCL pays predictable host launch overhead to avoid VCCL's unpredictable registration stalls.

\section{Conclusion}
This paper revisits a basic assumption in distributed training systems: collective communication is often optimized for bandwidth, while its consumption of GPU compute resources is overlooked. We present HOCCL to show that communication progress can be moved off GPU SMs and onto dedicated hardware, reducing interference between communication and computation. More broadly, HOCCL suggests that communication design should aim not only for high standalone speed, but also for preserving GPU resources for training. This perspective becomes increasingly important as modern training workloads grow larger and place ever greater pressure on limited GPU compute capacity. It also indicates that future communication libraries should be evaluated not only by their raw communication throughput, but also by how effectively they coexist with end-to-end training execution. Our results show that this perspective is both practical and beneficial, highlighting a promising direction for future large-scale training systems.
\bibliographystyle{plain}
\bibliography{refs}

\end{document}